\documentclass[12pt]{article}
\usepackage{epsf,latexsym}
\usepackage[round]{natbib}

\newcommand{\Tc}{T$_{c}$}
\newcommand{\Hcone}{H$_{c1}$}
\newcommand{\Hctwo}{H$_{c2}$}
\newcommand{\Hzero}{H$_{0}$}
\newcommand{\Hr}{H$_{r}$}

\newcommand{\etal}{{\em et al.}}

\begin{document}
\title{Magnetization Measurements on Single Crystals of Superconducting Ba$_{0.6}$K$%
_{0.4}$BiO$_{3}$}
\author{Donavan Hall\\
National High Magnetic Field Laboratory\\
Florida State University\\
Tallahassee, FL  32306-4005\\
\and
R. G. Goodrich and C. G. Grenier\\
Department of Physics and Astronomy\\
Louisiana State University, Baton Rouge, LA 70803-4001\\
\and
Pradeep Kumar\\
Department of Physics\\
University of Florida, Gainesville, FL 32611-8440\\
\and
Murali Chaparala\\
Department of Physics\\
University of Virginia, Charlottesville, VA 22901\\
\and
M. L. Norton\\
Department of Chemistry\\
Marshall University, Huntington, WV 25755-2520\\
}

\date{\today }
\maketitle

\begin{abstract}
Extensive measurements of the magnetization of superconducting single 
crystal samples of Ba$_{0.6}$K$_{0.4}$BiO$_{3}$\ have been made using 
SQUID and cantilever force magnetometry at temperatures ranging 
between 1.3 and 350 K and in magnetic fields from near zero to 27 T. 
Hysteresis curves of magnetization versus field allow a determination 
of the thermodynamic critical field, the reversibility field, and the 
upper critical field as a function of temperature.  The lower 
critical field is measured seperately and the Ginzburg-Landau 
parameter is found to be temperature dependent.  All critical fields 
have higher T = 0 limits than have been previously noted and none of 
the temperature dependence of the critical fields follow the expected 
power laws leading to possible alternate interpretation of the 
thermodynamic nature 
of the superconducting transition.
\end{abstract}

\vspace{0.5cm}

\begin{center}
	Accepted for publication in \emph{Philosophical Magazine B} on 7 August 1999
\end{center}

\section{Introduction}

The discovery of high T$_{c}$\ superconductivity in (La,Ba)$_{2}$CuO$_{4}$ 
\citep{Bednorz1986} was partly inspired by the bisthmuthate superconductor,
Ba(Pb, Bi)O$_{3}$.\citep*{Sleight1975} Subsequent cuprate superconductors
command the attention of a great many researchers because of their high T$
_{c}$s (on the order of 100 K). The discovery of high T$_{c}$\
superconductivity (HTSc) in the bismuthate system Ba$_{1-x}$K$_{x}$BiO$_{3}$%
\ (BKBO) more than ten years ago\citep*{Cava1988,Mattheiss1988b,Hinks1988a} 
with a T$_{c}$ of 30 K began an equally significant,
complementary course of research into the nature of HTSc for the 
following reasons: (1) it contains no copper, (2) it is an
isotropic conductor, (3) it has no evident magnetic ordering off 
stoichiometry, (4) its
superconductivity occurs at the boundary of a metal-insulator transition (a
commonality with the cuprates) \citep*{Hinks1988a,Dabrowski1988,Sato1989}, (5)
normal state resistivity measurements evince other than pure metallic behavior 
\citep*{Dabrowski1988}, (6) it has two related compounds: Ba$_{1-x}$Rb$_{x}$BiO%
$_{3}$, that has a T$_{c}$\ of $\approx$ 29 K for 0.28 $\leq$ x $\leq$ 0.44 
\citep*{Itti1991}, and Ba(Pb, Bi)O$_{3}$, a more conventional superconductor
with a T$_{c}$\ of $\approx$ 12 K \citep{Batlogg1984}, and (7) both BKBO and
Ba(Pb, Bi)0$_{3}$ have a comparatively low density of states (low carrier
densities), given their high values of T$_{c}$.

There have been several previous measurements of the magnetic
properties of superconducting Ba$_{1-x}$K$_{x}$BiO$_{3}$\ reported in
the literature \citep*{Batlogg1988b}, \citep*{Welp1988},
\citep*{McHenry1989}, \citep*{Grader1990}, \citep*{Huang1991}.  For x
= 0.4, single crystal measurements of H$_{c1}$\ were made by
\citet*{Grader1990} using an electrodynamic force balance technique on
micron-sized single crystal samples.  It was found that the lower
critical field values measured on their near perfect single crystals
were much higher (250 gauss at 10 K) than the H$_{c1}$ values reported
by \citet{Batlogg1988b} on powdered samples of similar composition (90
gauss at 10 K).  Further measurements of H$_{c1}$ on large single
crystals have been done by \citet{Huang1991} who also reported low
values of 95 gauss at 5 K. All of these measurements were done on
samples of imprecisely known demagnetization factors.  For single
crystals, in \citet{Grader1990}, H$_{c1}$\ is found to vary linearly
with temperature over the range of temperatures measured (7 - 22 K)
whereas in \citet{Huang1991} the H$_{c1}$\ versus temperature curve
shows a downward curvature as T = 0 is approached.

The upper critical field H$_{c2}$ also was reported in
\citet{Batlogg1988b}, and extensive H$_{c2}$ studies of powdered
samples of varying composition were done by \citet{Welp1988}.  In the
constant-field, temperature-dependent magnetic measurements done by
\citet{Batlogg1988b} a linear negative slope of dH$_{c2}$/dT = - 0.5
T/K between 20 and 29 K was observed.  Constant-field,
temperature-dependent resistivity measurements of several potassium
concentrations obtained by \citet{Welp1988}, show
that for x = 0.4, a slope of -0.6 T/K fit temperatures well
below T$_{c}$. A slight, positive curvature of dH$_{c2}$/dT was found near T$%
_{c}$ in all of the samples measured in \citet{Welp1988}.
Recently, \citet*{Samuely1996} have used tunneling
measurements on low T$_{c}$\ samples (T$_{c}$\ = 20 K) to determine H$_{c2}$%
\ as a function of temperature; they find much lower values of H$_{c2}$\
than in other measurements.

Several recent measurements of the magnetic properties of Ba$_{1-x}$K$_{x}$%
BiO$_{3}$\ in \citet*{Gatalskaya1996} and 
\citet*{Goll1996} and in an earlier paper by 
\citet*{McHenry1989} have focused on the reversibility field as a function of
temperature. In their study of flux creep in Ba$_{0.6}$K$_{0.4}$BiO$_{3}$,
\citet{McHenry1989}  found the reversibility field never to exceed 0.8 T
down to 5 K. In samples of nominal composition (x = 0.34), 
\citet{Gatalskaya1996} find good agreement near T$_{c}$\ using a
power law: H$_{r}$\ $\approx (1 - T/T_{c})^{m}$ with m = 1.45. For
samples having a T$_{c}$\ of 20 K, force magnetometer measurements over a
temperature range of 0.4 to 20 K by \citet{Goll1996} show
a continual upward curvature of H$_{r}$\ extending to 25 T at 0.4 K. Thus
different groups have measured widely different values for H$_{r}$\ on
different samples.

We report extensive studies of the magnetization of single crystal samples
of Ba$_{0.6}$K$_{0.4}$BiO$_{3}$. Magnetic field dependent hysteresis curves
of the magnetization at constant temperature were recorded at temperatures
ranging from 1.3 to 32 K in applied fields from zero to 27 T, and the
temperature dependence of the magnetization was measured from above room
temperature down to below the superconducting transition temperature. From
these measurements, we derive several quantities: (1) the complete
temperature dependence of the lower critical field H$_{c1}$, (2) the upper
critical field H$_{c2}$, (3) the reversibility field H$_{r}$, and (4) the
thermodynamic critical field H$_{0}$. All of these critical fields are
determined as a function of temperature. In addition, the temperature
dependence of the normal state susceptibility is determined.

These results highlight the unique nature of the properties of BKBO.
While the qualitative features of the magnetic properties are quite
reasonable, and look much like any other superconductor, in detail,
BKBO has a remarkable number of anomalous properties.  For example, in
a conventional superconductor, both \Hcone\ and \Hctwo\ have the same
temperature dependence.  Indeed the ratio, $H_{c2}/H_{c1}$, within
logarithmic factors, is the square of the Ginzburg-Landau (GL)
parameter, $\kappa$, and is temperature independent near \Tc.  In
BKBO, \Hcone\ and \Hctwo\ have different temperature dependences,
leading to a temperature dependent (and divergent at \Tc) $\kappa$. 
The thermodynamic critical field \Hzero\ $\propto (1-T/T_{c})^{2}$ in
contrast to a conventional superconductor where the exponent is one. 
The specific heat discontinuity at \Tc\ appears to be zero.  Likewise,
we find the discontinuity in the magnetic susceptibility to be zero. 
When all of these results are taken together, it has recently been
shown \citet*{Kumar1999} that a new model for the thermodynamic nature
of the phase transition from the normal to SC state gives good
agreement with the results.

This paper is organized as follows: In section \ref{sampprep} we 
briefly describe the sample preparation technique following Norton 
and Tang \citet*{Norton1991}.  Section \ref{measure} contains a 
description of measurement techniques in detail.  Section 
\ref{results}, divided into five subsections, goes through the 
results for \Hcone, \Hzero, \Hctwo, \Hr\ and finally the normal 
state, temperature dependent magnetic susceptibility.  Section 
\ref{discuss} presents our current understanding of these results, 
relating them to earlier measurements.  We also include a brief discussion 
of the properties of phase transitions of order $>$ 2, and how they 
relate to the results presented here.  The paper ends with a summary 
of our conclusions in section \ref{conclude}.

\section{Sample Preparation}
\label{sampprep}

The crystals used in this study were grown by scaling up the
electrosynthetic method described by \citet{Norton1991}. A
typical procedure, briefly listing the essential modifications, is given
here. An initial charge of 344 g of KOH (Baker, Reagent Grade) and 41 g Bi$%
_{2}$O$_{3}$ (Baker, Analyzed) was melted in a 300 ml PFA transfer container
(Berghof / America) using a temperature regulated (Omega) heating mantle
(Glas-Col), with stirring, producing a solution saturated with bismuth ions.
To this solution, 27 g of Ba(OH)$_{2}$. 8H$_{2}$O (Matheson, Reagent) was
slowly added, with stirring, under an atmosphere of flowing water saturated
nitrogen, an inert cover gas. After the solution clarified, and stabilized
at 240 $^{\circ}$C, a 2 mm diameter (0.1mm wall thickness) platinum tubing
cathode (Goodfellow, 99.95 \%), a 3.2 mm diameter bismuth rod reference
electrode (ESPI, 5N purity) and a 1 mm diameter silver wire anode (Aesar,
99.9 \%) were inserted 1 cm into the solution. A potential of 0.69 V was
applied utilizing a potentiostat (BAS model CV 27). Multiple crystals were
harvested after 340 hours by extraction of the silver anode with the firmly
attached crystals from the molten solution, followed by air cooling.
Adhering solidified flux was removed by rinsing with distilled water. The
samples were air dried, and the least flawed single crystals displaying
cubic morphology were chosen for this study.

The high T$_{c}$\ ($>$ 30 K) found for these crystals suggests that they
were very close to the optimal superconducting stoichiometry. Subsequent
X-ray analysis showed the 1 mm$^{3}$ facets used in these experiments were
single perovskite crystals with a mosaic spread of less than 2$^{\circ}$.

One of the problems in determining H$_{c1}$\ has been the fact that
the sample demagnetizing factor is of importance for this measurement. 
Without a controlled demagnetizing geometry, flux begins to penetrate
the sample at different fields depending on position in the sample. 
For this reason we produced a single crystal sphere for the H$_{c1}$\
measurements.  Starting with a crystal of approximately 1.5 x 1.5 x
1.5 mm$^{3}$ and a T$_{c}$\ of 30 K we ground it into a sphere in an
air driven racetrack using a diamond sandpaper abrasive.  The
resulting sphere had a diameter of 1 mm.  After grinding the sphere,
it was annealed in O$_{2}$ for 24 hours at 400 $^{\circ }$C. Laue
pictures showed no signs of strain.\footnote{We are indebted to Larry 
Hults and J. L. Smith of Los Alamos National Lab for annealing and 
x-raying this sample.} After this processing
the sample used for the H$_{c1}$\ measurements had a T$_{c}$\ of
approximately 29 K.

\section{Measurements}
\label{measure}
The measurements near T$_{c}$\ reported here were made using a Quantum
Design MPMS SQUID magnetometer (B $<$ 5.5 T) .  Several precautions
must be made in measurements on superconductors near T$_{c}$\ using a
measurement system of this type.  There are two basic problems. 
First, the magnetic moment of the sample holder can easily become
larger than that of the sample near T$_{c}$, and the maximum moment of
the sample holder combined with the sample shifts along an axis
parallel to the applied field.  When this occurs, the output voltage
from the SQUID pickup coils is not centered and spurious results in
the calculation of the magnetic moment can occur.  To overcome this
problem, we held the samples in long quartz tubes having an inner
diameter slightly smaller than the largest dimension of the sample to
be measured.  The empty quartz tube extends through all three pickup
coils of the magnetometer during a complete vertical scan of the
magnetometer.  It was verified that the tubes produced a negligible
signal over the scan length used.  For most of the SQUID
measurements, the samples were inserted into the tubes with the
maximum dimension parallel to the axis of the tube, then turned
slightly to hold them in place.  Care was taken to center the sample
in the pickup coils at each temperature because of the displacement
due to thermal contraction of the quartz tube.

When the measurements
on the spherical sample were done, the above procedure was not
possible and it was secured to the inside of the quartz tube with a
small piece of Kapton tape.  This mounting limited the measurements to
a minimum field of about 0.5 mT. The second problem with low field
SQUID magnetometry is that the superconducting magnet retains trapped
flux at low fields.  The trapped flux can be removed partially by
cycling between positive and negative fields with decreasing final
fields.  This was done at each temperature at which a measurement for
H$_{c1}$\ was performed.

For temperatures between 2 and 32 K both constant temperature magnetic field
hysteresis curves and constant field temperature hysteresis curves were
recorded on two different crystals. At constant temperatures, measurements
were taken as the field was raised from zero to a positive value greater
than H$_{c2}$, then lowered through zero to a negative field value greater
than H$_{c2}$. It was noted that any subsequent cycling of the field does
not supply any additional information concerning the critical points. Hence,
the symmetry of the complete hysteresis curves could be used to reduce the
number of field points taken at a given temperature. An example of a
complete constant temperature (T = 22.5 K) hysteresis curve is shown in
Figure \ref{f1cthyst} with an expanded view showing H$_{r}$\ and H$_{c2}$\
in the inset. The temperature of the sample was raised to 50
K (T $>$ T$_{c}$ at H = 0) before each constant temperature hysteresis
measurement, then lowered to the measurement temperature in zero field.
Thus, all constant temperature data reported are for zero field cooled
samples. Samples not cooled in zero field would have a non-zero
magnetization at zero field, but the magnetization curve for increasing
field merges with the decreasing field portion at the same value of H$_{r}$.

Further measurements of H$_{c2}$\ and H$_{r}$\ were done using a cantilever
force magnetometer at the National High Magnetic Field Laboratory (NHMFL) in
a 27 T resistive magnet at temperatures from 1.4 to 19 K with the 
applied magnetic field being swept at a rate of 0.25 T/minute.  The sample used
for these measurements had a T$_{c}$\ of 30 K. An example of the constant
temperature magnetic hysteresis curve taken with the force magnetometer is
provided in Figure \ref{f2cthyst}.

The constant field measurements were started at a temperature above T$_{c}$.
The samples were cooled to 4.5 K in zero field, the field applied, the
temperature slowly raised, point by point, to above T$_{c}$ while recording
magnetization, then again cooled to 4.5 K during the measurement in the
applied field. An example of a complete cycle of this type of data is shown
in Figure \ref{chmvstemp}. Again, near T$_{c}$ the magnetization is
extremely small and care was taken to avoid the effects of the sample
holder. Finally, we measured the magnetization at several temperatures 
at two
constant fields in the normal state from 32 to 300 K to determine the normal
state susceptibility.

\section{Results}
\label{results}
In addition to the separate H$_{c1}$\ measurements, there are several
critical fields to be obtained from the hysteresis data: H$_{0}$, H$_{c2}$,
and H$_{r}$. The thermodynamic critical field H$_{0}$\ is obtained only from
the constant temperature data while H$_{c2}$\ and H$_{r}$\ were measured
both at constant temperature and constant field. None of the critical fields
can be determined in a straightforward manner and we state in detail how we
have extracted them from the data. Between approximately 18 and 32 K, the
two highest critical fields were determined using the SQUID magnetometer.
Below 18 K, H$_{c2}$\ was in excess of the maximum field of the SQUID
magnetometer (5.5 T). Above 22 K, H$_{c1}$\ was sufficiently low that no
reliable data could be obtained.

\subsection{H$_{c1}$}

Values of H$_{c1}$\ reported here were obtained from the spherical sample
from 4.2 to approximately 22 K. Typical low temperature, low field
magnetization curves for several temperatures are shown in Figure \ref
{ssmvshpolytemp}. The initial slope of the curve is given by M/H = V/4$\pi $
(1-D), where V is the volume of the sample and D is the demagnetization
factor. Only when the sample geometry is such that D = 0 will this constant
slope extend to H$_{c1}$, at which point the magnetization would abruptly
decrease. When D $\neq $ 0, the field begins to penetrate the sample at
different points gradually until the entire sample is in the mixed state. We
shaped a single crystal sample into a sphere, (D = 1/3), hence the initial
slope is much larger and the transition from the Meissner to the mixed state
much sharper than for the crystals that have non-symmetric shapes.

To determine the value of H$_{c1}$, the following procedure was used. The
low field, linear portion of the curve was fit with a straight line, 
and the
field value at which the measured magnetization deviated from the
straight line was taken to be $\frac{2}{3}$H$_{c1}$. Values obtained from
this fit for H$_{c1}$\ are plotted as a function of temperature in Figure 
\ref{sshc1vst}.  Note that below 15 K these results give the same 
values as those reported in \citet{Grader1990}.

\subsection{H$_{0}$}

The thermodynamic critical field is obtained by integrating over half the
cycle, M$_{R}$(H) = [M(H$_{incr}$) + M(H$_{decr}$)]/2, from zero to $>$ H$%
_{c2}$. The value of H$_{0}$\ is obtained from: 
\begin{equation}
\int_{}^{}M\cdot dH = \frac{H_{0}^{2}}{8 \pi}.  \label{hzerointegrate}
\end{equation}
Since SQUID magnetometry measurements require a constant applied field
during the measurement, and the data were accumulated point by point we have
numerically integrated the hysteresis curves to obtain H$_{0}$ at each
temperature. The temperature dependence of H$_{0}$\ is shown in Figure \ref
{h0vst} for the sample used in the SQUID measurements. We have elected not
to calculate values for H$_{0}$\ from the high field-low temperature data
until the anomalous ``fishtail'' structure (see below), also noticed by
\citet{Gatalskaya1996}, is fully understood. The
fishtail adds area below the magnetization curve; thus, thwarting any
attempt of straightforward analysis.

\subsection{H$_{c2}$}

A typical plot of the magnetization for
increasing field from zero tesla to above H$_{c2}$\ is shown in Figure \ref
{f1cthyst}. There are several points about this curve that make
determination of H$_{c2}$\ complicated. Between H$_{c1}$\ and H$_{c2}$\ the
curve is never linear in applied field, and an extrapolation to H$_{c2}$\
from a linear portion of the curve in the superconducting state near H$_{c2}$%
\ is not possible \citep*{Hao1991}. Above H$_{c2}$ BKBO is diamagnetic (mainly
due to the atomic core contribution to the magnetization) and the total
magnetization never becomes positive upon passing from the superconducting
to the normal state. To obtain a consistent value for H$_{c2}$, we have fit
the linear diamagnetism data in the normal state above H$_{c2}$ to a
straight line and taken the field at which the magnetization deviates from
the extrapolation of this line to zero field to be H$_{c2}$. From values of H%
$_{c2}$\ obtained in this manner, we show the temperature dependence of H$
_{c2}$\ in Figure \ref{hc2vst} along with the
determination of H$_{c2}$\ from the constant field data to be discussed
below. The minimum in the magnetization curve in this field region is fairly
broad for all of the samples, but this technique gives values of H$_{c2}$\
consistent with the ones that are obtained from temperature dependent data
at constant field \citep{Batlogg1988b,Huang1991}.

We observe the anomalous ``fishtail'' structure in our force magnetometer
measurements also reported by \citet{Gatalskaya1996}.
These structures are not observed in the magnetization data taken with the
SQUID magnetometer. This leads us to question the origin of these fishtails,
specifically, whether they might be due to the measurement technique. The
fishtails seen in our force magnetometer measurements might be due to a
magnetic field sweep rate that exceeds the flux lattice relaxation time, or
may be associated with uncompensated torque contributions from the sample.
It should be noted that these fishtail structures do not affect our ability
to determine H$_{r}$\ and H$_{c2}$\ from these data, and these two 
critical fields are the only
information we have extracted from the force measurements.

Using the 21 K values of H$_{c1}$\ and H$_{c2}$, the
GL parameter, $\kappa$, was calculated from the relation
\begin{equation}
\frac{ln\kappa}{2\kappa^{2}} = \frac{H_{c1}}{H_{c2}}.  \label{ginz-land}
\end{equation}
We find that the value of $\kappa$ at 21 K is approximately 64. This compares
favorably with the value reported by \citet*{Kwok1989} of $%
\kappa$ = 59.

\subsection{H$_{r}$}

At temperatures above 15 K, we are able to determine
the value of applied field at which the hysteretic behavior becomes
reversible, H$_{r}$. We have taken this field to be the lowest value of H at
which the increasing and decreasing applied field have the same measured
values of M. The temperature dependence of H$_{r}$ is shown in
Figure \ref{hc2vst}.

\subsection{Temperature Dependent Constant Field Data}

Measurements of the temperature dependence of the magnetization at constant
field give the same information in the T-H plane through measurements of
critical temperatures, T$_{c2}$\ and T$_{r}$, as do the constant temperature
field dependent measurements. A view of one set of data taken at constant
field is shown in Figure \ref{chmvstemp}. The temperature at which the
magnetization deviates from a constant on the high temperature side is taken
to be T$_{c2}$. The temperature at which the increasing and decreasing
measurements deviate is T$_{r}$. One can see from the data presented 
in the inset of
Figure \ref{hc2vst} that the two different measurements give the same
result. This agreement shows that the phase boundaries between the normal and
superconducting states and between the reversible and irreversible flux
regions within the superconducting states are equilibrium thermodynamic
boundaries and are not controlled by non-equilibrium phenomenon.

\subsection{Normal State Susceptibility}

The normal state susceptibility, $\chi$, was calculated from the difference of two
magnetization curves taken at 0.3 T and 5.0 T at temperatures from 31 K to
350 K and is shown in Figure \ref{normmag} with $\chi $ as a function of 
1/T in
the inset. The difference between the magnetization measurements, $\Delta $M,
divided by the field difference, $\Delta $H, was taken to be the
susceptibility, $\chi =\Delta M/\Delta H$. The reason for the large field
difference for the two magnetization measurements is that M is small in each
case, and the large $\Delta $H reduces errors in $\Delta $M.  As long 
as the magnetization remains in the linear response regime, the 
relatively large size of $\Delta$H is of no concern.  The quartz
tube sample holder contributes a small signal to the magnetization; this has
been measured and subtracted from the difference magnetization.

\section{Discussion}
\label{discuss}
The temperature dependence and the magnitudes of the critical fields
measured here are different from that expected from an ordinary type-II
superconductor. In general, the critical fields have different 
temperature dependences
from what is expected and some new phenomena are observed. We begin with a
discussion of the normal state susceptibility.

In their initial measurements of the normal state susceptibility, 
\citet{Cava1988} found large extraneous paramagnetic contributions
presumedly due to the presence of unreacted KO$_{2}$ in their samples. In a
following paper \citep{Batlogg1988b}, the same authors corrected their data
with an estimate of the core diamagnetism ($\chi _{c}\approx $ $-7.5\times %
10^{-5}$ emu/mol) which is the same value used here.  From their
energy band calculations
\citet*{Mattheiss1988a} determined the density of states at E$_{F}$ in Ba$%
_{0.6}$K$_{0.4}$BiO$_{3}$\ to be 0.46 states/eV. Using this value, we
calculate the theoretical contribution to the susceptibility due to
Pauli paramagnetism to be $2.7\times 10^{-5}$ emu/mol.  Our
measurements show that the normal state susceptibility at T = 300 K is $-5.3%
\times 10^{-5}$ emu/mol (see Figure \ref{normmag}), and this value agrees
with that reported by \citet{Hundley1989} and 
\citet{Uwe1996} for a potassium concentration of x = 0.40. Our
measurements show an increase in the normal state paramagnetism below 150 K.
We have fit the data below 150 K to a Curie-Weiss function plus
a temperature independent Pauli term, 
\begin{equation}
(\chi -\chi _{core})=\chi _{P}+\frac{C}{T-\Theta },  \label{curie-weiss}
\end{equation}
where $\chi _{core}$ is the core diamagnetism, $\chi _{P}$ is the Pauli term, 
$C$ is the Curie constant, and $\Theta $ is the ordering temperature. The
values obtained from the fits are $\chi _{P}$ = 1.81 $\times $ 10$^{-5}$
emu/mol, $C$ = 8.87 $\times $ 10$^{-4}$ erg K/G, and $\Theta $ = 2.48 K for
one sample and $\chi _{P}$ = 1.81 $\times $ 10$^{-5}$ emu/mol, $C$ = 8.28 $%
\times $ 10$^{-4}$ erg K/G, and $\Theta $ = 2.21 K for a second sample. Using the
values of $C$ and assuming the impurities have a spin quantum number of $%
\frac{1}{2}$, the magnetic impurity concentration is estimated to be
0.66 \% and 0.71 \% for the two samples.  If the
impurities have a larger spin quantum number, then the magnetic
impurity concentrations would be lower.  The Pauli term for both
samples is the same and is approximately 2/3 of that
obtained from band theory \citep{Mattheiss1988a}. We point out that we have interpreted the linear 1 /%
$\chi $ vs. T behavior at low temperatures as being due to localized spin impurities. However,
this behavior might also arise from other mechanisms intrinsic to the normal
state.

If an overall linear fit to the H$_{c1}$ data is done, a slope of dH$_{c1}$
/dT = -17.2 $\pm $ 0.66 G/K is obtained as compared to -11 G/K from 
\citet{Grader1990}. These values are much larger than the value
of \citet{Batlogg1988b} of -4.5 $\pm $ 0.5 G/K; 
also, the
values of H$_{c1}$ derived from single crystals are larger than the 
powder samples used by Batlogg \etal. From this linear fit to the H$_{c1}$%
\ data, we find that H$_{c1}$(T=0) is 393 $\pm $ 6 G. However, it is clear
that a linear function does not  represent best the H$_{c1}$\ dependence on
temperature. Between 5 and 15 K the data appears to be linear.
Around 15 K, H$_{c1}$(T) experiences an inflection point and has a 
different
curvature at higher temperatures. Attempts were made to measure H$_{c1}$\ at
fields approaching T$_{c}$, but above 22 K the magnetization signal was too
small to resolve with the SQUID measurement system near zero applied 
field.

We find that a good fit to the H$_{c1}$ data above 12 K is obtained 
from H$_{c1} (T) \propto (1 - T/T_{c})^{3}$ with the curve passing 
through the H = 0 value of T$_{c}$ = 29 K.  The very low value of H$%
_{c1}$\ is the reason the very broad temperature dependent magnetic
transitions are observed in BKBO, even though the zero field resistive
transitions are sharp. All of the magnetic measurements are made in small
magnetic fields and continuous changes in M(T) are observed until the
temperature is reached where H$_{c1}$ is greater than the measurement
field. We note that H$_{c1}$\ is the only measured thermodynamic critical
field that does not have upward curvature as the temperature goes to zero.

As a result of our upper critical field measurements, we found that a
power law fit where H$_{c2}$\ and H$_{r}$\ $\sim $ (1 -
T/T$_{c}$)$^{m}$ yielded values of m = 1.58 and 1.97 for H$_{c2}$\ and
H$_{r}$\ respectively (as shown in Figure \ref{fmlnlnhc2andhr}). 
\citet{Gatalskaya1996} found for H$_{r}$\ that m = 1.45 for a sample
with a potassium concentration of 0.34.  Also, \citet{Goll1996} report
a value of m = 1.5 for H$_{r}$\ in their below optimum potassium
concentration sample.  Our samples had a nominal potassium
concentration of 0.40.

\citet{Welp1988} report upper critical field slopes for several
concentrations of K, but no positive curvature of the H$_{c2}$\ versus
temperature was reported.  Like \citet{Affronte1994} and
\citet{Gantmakher1996}, we observe a positive curvature in the H$%
_{c2}$\ versus temperature curve; however, we observe an enhanced curvature
as compared to the results of \citet{Gantmakher1996}.
This curvature deviates significantly from the universal behavior predicted
by Werthamer, Helfand, and Hohenberg (WHH) \citep*{Werthamer1966} for
superconductors with weak electron-phonon coupling. 
\citet{Gantmakher1996} find empirically that their H$_{c2}$\ data fits the
function 
\begin{equation}
H_{c2}(T)=32.2-1.8\: T+0.025\:T^{2}.  \label{gantmakher1}
\end{equation}
A plot of this equation is shown in Figure \ref{hc2vst}. Our H$_{c2}$\ data show a greater curvature
than the fit of \citet{Gantmakher1996} and it
appears a second order fit greatly underestimates the probable value of H$%
_{c2}$(T=0), and their value of 32.2 T might be too low. 
\citet{Gantmakher1996} comment that the positive curvature is enhanced by
disorder in the sample; therefore, the enhancement of the curvature and
correspondingly in H$_{c2}$(T=0) may be due to an intrinsic disorder in BKBO.

From our determination of H$_{r}$\ we observe that the irreversibility 
line also
deviates from WHH theory and displays a positive curvature. Similarly, 
\citet{Goll1996} observe a positive curvature in H$_{r}$\
versus temperature, and again our values of H$_{r}$\ are enhanced when
compared with the values of \citet{Goll1996}.  Similarly, the thermodynamic
critical field, H$_{0}$, has a positive curvature as is shown in Figure \ref
{h0vst}.

\citet{Samuely1996} have determined the upper critical
field H$_{c2}$\ by Andreev reflection in point-contact junctions, a
measurement of the superconducting density of states. They find no positive
curvature of the H$_{c2}$\ versus temperature curve and claim that their
measurements confirm an adherence of BKBO to WHH theory. This contradicts
all other magneto-transport measurements \citep*{Affronte1994} that uniformly
show H$_{c2}$\ and H$_{r}$\ as a function of temperature in BKBO deviate
from WHH theory. The only remarkable difference between the samples used by
\citet{Samuely1996} and other groups is that their samples had a T$_{c}$\
of $\approx $ 23 K. Studies, including this one, that use samples with 
T$_{c}$s of 30 to 32 K generally show a deviation from the WHH theory.

Measurement of magnetization across a phase transition may contain
fluctuation effects. Following early works of \citet*{Aslamazov1968},
\citet*{Ullah1991} have proposed a scaling
equation for temperature and field dependence of magnetization. The scaling
variables used are 
\begin{eqnarray}
\left[ T-T_{c}(H)\right] /[HT]^{2/3}\quad \quad \quad D &=&3  \nonumber \\
\left[ T-T_{c}(H)\right] /[HT]^{1/2}\quad \quad \quad D &=&2
\end{eqnarray}
where the dimensionality of the electron system is given by D. For either
D=3 or D=2 the magnetization is expected to follow a universal curve. The
essential feature is that a smooth cross over takes place from a normal
paramagnetic (or diamagnetic in the case of BKBO) state to the mixed state
of a superconductor. Thermodynamic
quantities such as the magnetization and specific heat for YBCO have been
shown to fit a 3D universal curve by \citet{Welp1991} The
results of our H$_{c2}$ data being analyzed in this manner are shown in
Figure \ref{scaling}. One can see that the data for magnetization collapse onto a
universal curve. However, given that BKBO is a cubic isotropic system, it is
surprising that our results are just as well described by D=2 or D=3. It
appears that fluctuation theory cannot distinguish between 2 and 3
dimensional behavior in this case. Therefore, the broadening of the
transition may be due to effects other than fluctuations.

Several studies of the specific heat of BKBO have been done 
\citep*{Hundley1989,Graebner1989,Stupp1989,Alba1992}. All but one study 
\citep{Graebner1989}) conclude that $\Delta C_{p}$ is zero across the
transition. Because of the large phonon contribution to the specific heat
and the low density of electronic states, it is difficult to extract the
comparatively small value of the electronic specific heat change across the
transition. A calculation by \citet{Kwok1989} shows the BCS
discontinuity in BKBO to be $\Delta $C$_{p}$/T$_{c}$\ $\approx $ 3.75
mJ/mole K$^{2}$ or $\Delta $C$_{p}$ = 120 mJ/mole K if T$_{c}$\ = 32 K. This
value is calculated from the relation 
\begin{equation}
\Delta C_{p}/T_{c}=\left( \frac{1}{8 \pi \kappa ^{2}}\right) \left( \frac{%
dH_{c2}}{dT}\right) ^{2},  \label{deltaC}
\end{equation}
where $\kappa \simeq 64$ is the Ginzburg-Landau parameter at low 
temperatures.  At temperatures near T$_{c}$%
, the magnitude of the total specific heat is in the range of 15 to 20
J/mole K, and the expected discontinuity is approximately 0.1 J/mole K.
Previous measurements of the specific heat performed at LSU \citep*{XuUnpub}
using a sensitive ac technique \citep*{Xu1990} did not detect a discontinuity
in $C_{p}$ at the transition temperature and we believe the measurements
were of sufficient accuracy to detect a $\Delta C_{p}$ of the order of
magnitude predicted by \citet{Kwok1989}. Recently, more
high sensitivity measurements have been performed by Kim and Stewart 
\footnote{J. Kim and G. Stewart, University of Florida, Private 
Communication.} Again no discontinuity in $C_{p}$ is observed.

If one observes a second order phase transition as the 
normal-superconducting (N-S) phase boundary is crossed, a signature 
of the transition can 
be obtained from the temperature dependence of the thermodynamic critical
field. In a second order phase transition, the free energy $F=H_{0}^{2}/8\pi 
\propto (1-T/T_{c})^{2}$. Thus, $H_{0}(T)$ is expected to be linear in $%
(1-T/T_{c})$. In Figure \ref{h0vst}, the measured temperature
dependence of the thermodynamic critical field H$_{0}$ is given with the fit
to $(1-T/T_{c})^{2}$shown in the inset. This result clearly shows that 
the usual second order phase theory does not fit the data.

The experimental results for critical fields 
\Hcone, \Hr, and \Hctwo\ also are different from 
conventional superconductors in detail.  More specifically, they are 
anomalous in subtle but important ways.

First, the GL parameter $\kappa$ as seen from equation \ref{ginz-land}
is temperature dependent (see Figure \ref{kappavst}).  In a BCS-GL
framework \citep{Tinkham}, $\kappa$ ($= \lambda/\xi$) is temperature
independent near \Tc.  Indeed the entire classification of
superconductivity between types I and II rests on a constant $\kappa$
(whether $< 1/\sqrt{2}$ or not).

We see here that $H_{c1} \propto (1-T/T_{c})^{3}$ and $H_{c2} \propto  
(1-T/T_{c})^{3/2}$, leading to $\kappa \propto (1-T/T_{c})^{-3/4}$ 
using equation \ref{ginz-land}.  The divergence of $\kappa$ near 
\Tc\ indicates a much softer Meissner effect than seen in 
conventional superconductors.

Next, the 
principal premise of BCS-GL theory is a second order phase 
transition.  The Ehrenfest relation for this second order phase 
transition is given by
\begin{equation}
	\left(\frac{dH_{c2}}{dT}\right)^{2} = \frac{\Delta C}{T_{c}\Delta 
	\chi},
	\label{eq:phaseboundary1}
\end{equation}
where $\Delta C$ is the discontinuity in the specific heat and 
$\Delta \chi$ is the discontinuity in the magnetic susceptibility.  As 
stated above, there is evidence that $\Delta C = 0$.  In our 
measurements, we find $\Delta \chi = 0$.

It is worth noting that BKBO, a diamagnet, is a propitious 
superconductor to look for non-BCS-GL effects.  In a 
conventional superconductor, in the mixed state, the magnetic response 
is diamagnetic, although the magnetic susceptibility ($\frac{\partial 
M}{\partial H}$) is positive, since M is negative.  However, in the normal state most 
metals are weakly paramagnetic.  Thus when the transition 
from a superconductor to a normal metal occurs, $\Delta M \neq 0$ and 
the transition is weakly first order.  In BKBO, there is no 
discontinuity in M; the magnetization smoothly connects to the 
diamagetism of the normal state.  Since both $\Delta M$ and $\Delta 
\chi$ appear to be zero, the order of the phase transition could be higher 
than two.  We point out that an indication of the transition order can be obtained from 
the thermodynamic critical field \Hzero.  In a phase transition of 
order n, $F = \frac{H_{0}^{2}}{8\pi} \propto (1-T/T_{c})^{n}$.  Thus the fit 
shown in Figure \ref{h0vst} suggests a fourth order phase transition 
where $H_{0} \propto (1-T/T_{c})^{2}$.

If the transition is indeed higher than second order in the Ehrenfest sense, 
then, the temperature dependence of 
specific heat will not show a discontinuity at T$_{c}$.
There should be a discontinuity in the derivatives of the specific heat; 
however, a measurement near \Tc\ will show effectively no sharp change in 
C.

In a previous paper \citep*{Kumar1997}, the properties of a third order
phase transtion have been developed.  When these same arguments are
extended to a fourth order phase transition \citep{Kumar1999} with a
free energy F$_{IV}$ the temperature dependance of H$_{c1}$ and
H$_{c2}$ can be calculated from the free energy F$_{IV}$ and the
length scale $\xi^{2} \sim c/a \sim (1 - T/T_{c})^{-1}$ yeilding
$H_{c2} = \Phi_{0}/2\pi \xi^{2} \propto (1 - T/T_{c})$.  The exponent
is obtained in a mean field analysis and is lower than the observed
exponent of 1.5 (1.25 near T$_{c}$, see Fig.  \ref{fmlnlnhc2andhr}
caption).  However, as pointed out by 
\citet{Gantmakher1996} the curvature in H$_{c2}$ can be enhanced by sample
disorder.  Next, the temperature dependence of $H_{c1} \sim
\phi_{0}/\lambda^{2} \propto (1 - T/T_{c})^{3}$.  This is in agreement
with the experiment; as is the corresponding $\kappa = \lambda/\xi$ in
its temperature dependence.

Our conclusions about the possible order of the SC transition are 
published in \citet{Kumar1999}; however  
much work still needs to be done to properly understand the nature of 
inhomogeneities and the effects they have on the critical magnetic fields.  
In addition, further analysis of fluctuations needs to be done.  Thus, 
whether macroscopic fluctuations can lead to a different temperature 
dependance of the irreversibility 
field \Hr\ which can influence the calculated values of \Hzero\ also 
remains to be analyzed.

\section{Conclusions}
\label{conclude}
Our magnetization measurements have been carried out as field
dependence at constant temperature and temperature dependence at
constant field to determine the critical fields for BKBO. In all cases
the agreement between the results of these two independent
measurements indicate that we are measuring equilibrium properties.

From high temperature (31 - 350 K) susceptibility measurements we determine
the conduction electron Pauli susceptibility to be 1.8 $\times $ 10$^{-5}$
emu/mol for two samples.  The total susceptibility also contains an 
ostensible ferromagnetic term with a Curie-Weiss temperature near 2 K.

We report results of the first force magnetometer measurements of BKBO to
fields up to 27 T, and show that the H$_{c2}$\ and H$_{r}$\ critical field
curves deviate from WHH theory \citep{Werthamer1966}, displaying a positive
curvature at low temperatures. Our overall results including both force and
SQUID measurements for these two critical fields are in general agreement
with previous measurements, but show higher values of H$_{c2}$\ and H$_{r}$
at low temperatures than previously have been reported. We find that while
fluctuation scaling theory fits the data from the SQUID measurements near $%
H_{c2}$, no distinction can be made between two or three dimensional theory.
This result indicates that fluctuations may not be the cause of the
broadness of the transition.

Finally, we note that BKBO may exhibit a phase transition that is not 
of 
thermodynamic order two.  The conclusion that the order is higher 
than two is based on previous measurements of the specific heat in the 
normal and superconducting state across the transition (no observed 
discontinuity), on 
measurements of the susceptibility across the transition (no 
$\Delta\chi$), and on the 
fact that BKBO is a normal state diamagnet.  The temperature 
dependence of the thermodynamic critical field (near \Tc), has led us 
to suggest that the transition is of order four.\citep{Kumar1999}  Furthermore, this is 
consistent with the experimental temperature dependence of \Hcone\ and 
is close to the temperature dependence of \Hctwo.  We note that near \Tc, the 
Ginzburg-Landau parameter, $\kappa = \frac{\lambda}{\xi}$, is 
temperature dependent, $\kappa \sim (1 - T/T_{c})^{-3/4}$, a 
divergence which points to an unusually soft Meissner effect at \Tc.

Perhaps the most significant difficulty in analysis of the present
data (and the data in earlier references on thermodynamic properties
of BKBO) is that one's intuition is based on a tacit assumption that
all superconducting transitions are second order with associated
fluctuation effects.  The expected properties of higher order phase
transtions are different yet one tries to understand an anomalous
property in terms which implicity assume a second order phase
transition.  However, the evidence presented here for a higher order
phase transition is by no means conclusive.

The experimental data presented here is (1) largely consistent with
earlier efforts on samples of similar quality and (2) where the
results are new, there may be a formalism, based on a higher order
phase transition that may explain the known
anomalies.

\section{Acknowledgments}
The work at LSU was supported by the NSF under grant No. DMR-9501419. A
portion of this work was performed at the National High Magnetic Field
Laboratory, which is supported by NSF Cooperative Agreement No. DMR-9016241
and by the State of Florida. The Norton group effort was supported by the
NSF under grant No. CHE-9612568.



\newpage
\begin{center}
	FIGURE CAPTIONS
\end{center}

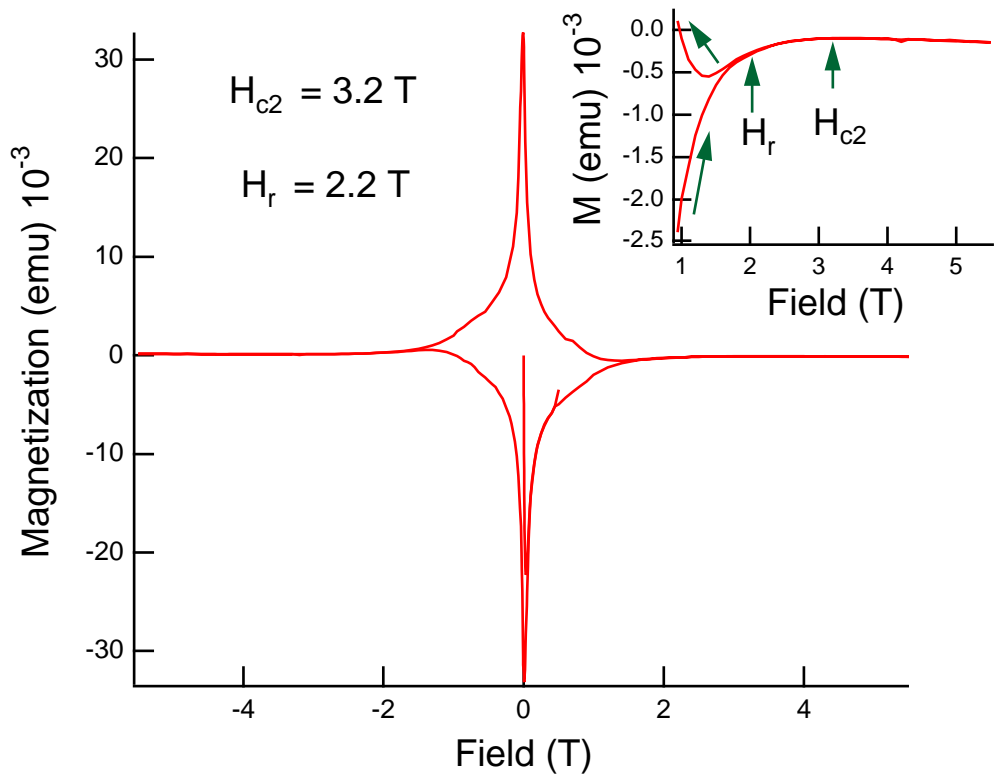
\begin{figure}[tbp]
	$$\epsffile{maghyst_inset.epsf}$$
	\caption{}
\label{f1cthyst}
\end{figure}

Figure \ref{f1cthyst}.
Complete hysteresis curve taken at 22. 5 K from H = 0 to 5.5 
T, back through zero to - 5.5 T and then to $>$ zero.  An expanded 
view of the data showing \Hr\ and \Hctwo\ is given in the inset.

\vspace{0.25cm}

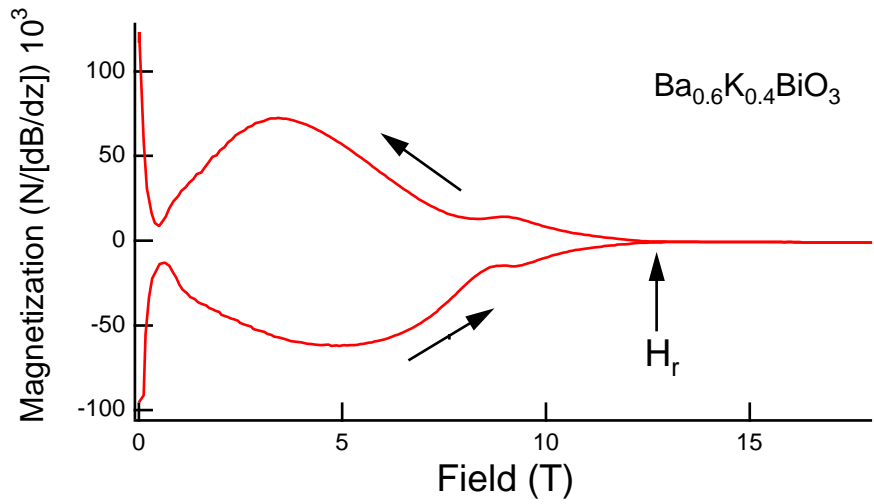
\begin{figure}[tbp]
	$$\epsffile{f2cthyst.epsf}$$
	\caption{}
\label{f2cthyst}
\end{figure}

Figure \ref{f2cthyst}.
Hysteresis in the magnetization as a function of applied field of Ba%
$_{0.6}$K$_{0.4}$BiO$_{3}$\ at 6.0 K from zero to $>$ +H$_{c2}$\ to zero.
This data is the magnetization determined from the force magnetometer. H$_{r}
$\ is indicated. The arrows show the direction of the field sweep. Note that
as H goes to zero, the field gradient, and therefore, the force also goes to
zero.

\vspace{0.25cm}

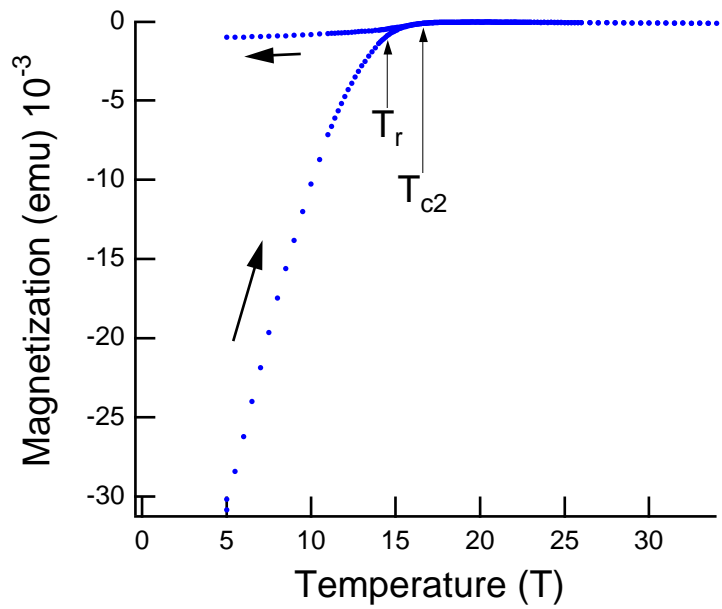
\begin{figure}[tbp]
	$$\epsffile{chmvstemp.epsf}$$
	\caption{}
\label{chmvstemp}
\end{figure}

Figure \ref{chmvstemp}.
Example of a magnetization hysteresis taken at constant 
field (H = 50000 G).

\vspace{0.25cm}

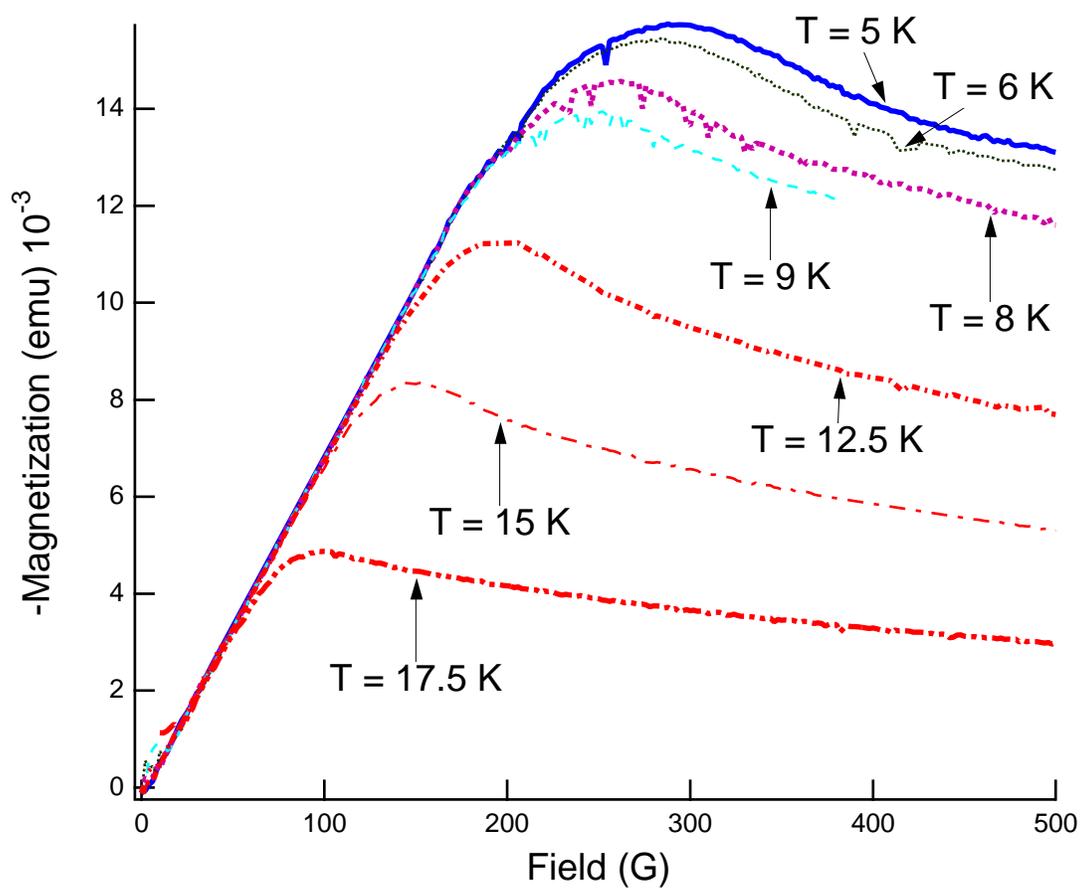
\begin{figure}[tbp]
	$$\epsffile{ssmvshpolytemp.epsf}$$
	\caption{}
\label{ssmvshpolytemp}
\end{figure}

Figure \ref{ssmvshpolytemp}.
Magnetization as a function of applied field for the spherical
sample of Ba$_{0.6}$K$_{0.4}$BiO$_{3}$\ at a number of temperatures. The
temperatures from from 5 K for the highest curve to 17.5 K for the lowest.

\vspace{0.25cm}

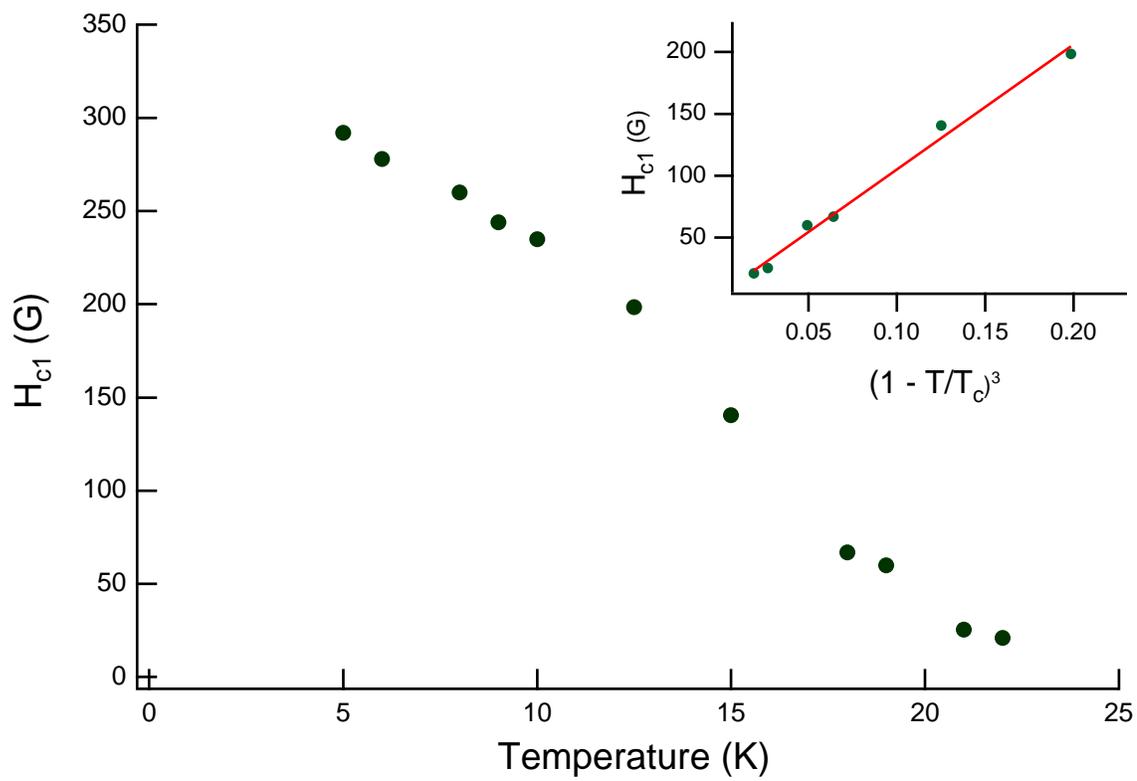
\begin{figure}[tbp]
	$$\epsffile{sshc1vst.epsf}$$
	\caption{}
\label{sshc1vst}
\end{figure}

Figure \ref{sshc1vst}.
Values of H$_{c1}$\ of Ba$_{0.6}$K$_{0.4}$BiO$_{3}$\ as a function
of temperature determined from magnetization data on the spherical 
sample.  The inset shows that \Hcone\ goes as $(1 - T/T_{c})^{3}$ for 
temperatures higher than 12 K.

\vspace{0.25cm}

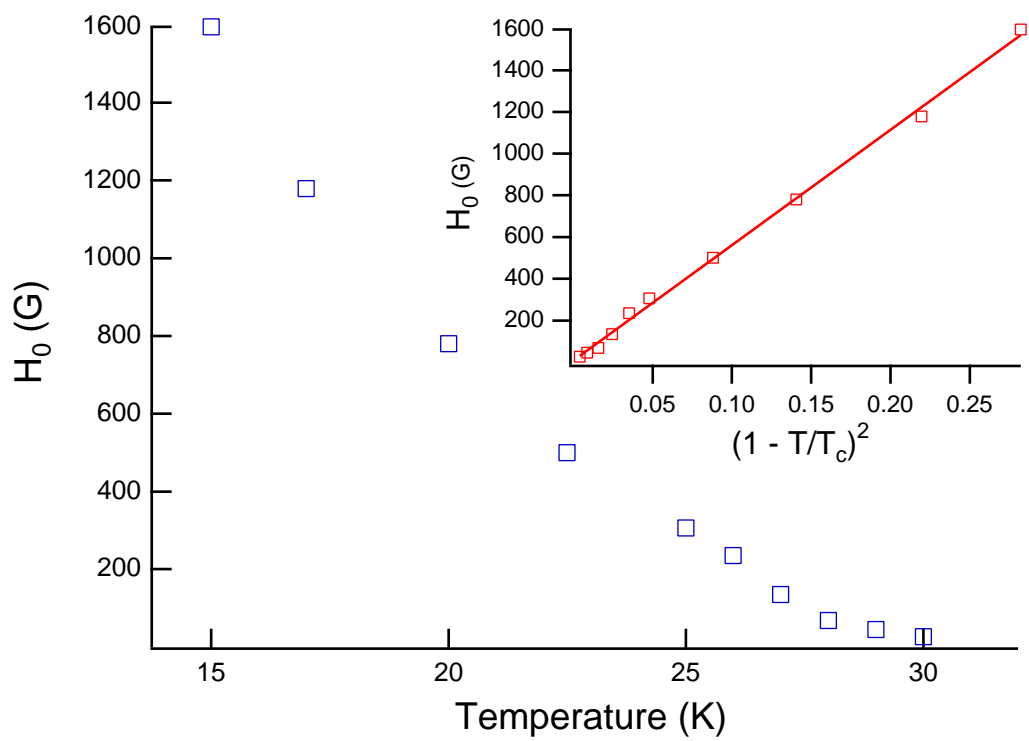
\begin{figure}[tbp]
	$$\epsffile{hc0vst.epsf}$$
	\caption{}
\label{h0vst}
\end{figure}

Figure \ref{h0vst}.
Values of the thermodynamic critical field for BKBO obtained
from integration of the magnetic hysteresis curves as a function of
temperature. The inset shows that in the fit $(1 - T/T_{c})^{n/2}$ that n =
4, thereby showing that the superconducting transition is of order 4.

\vspace{0.25cm}

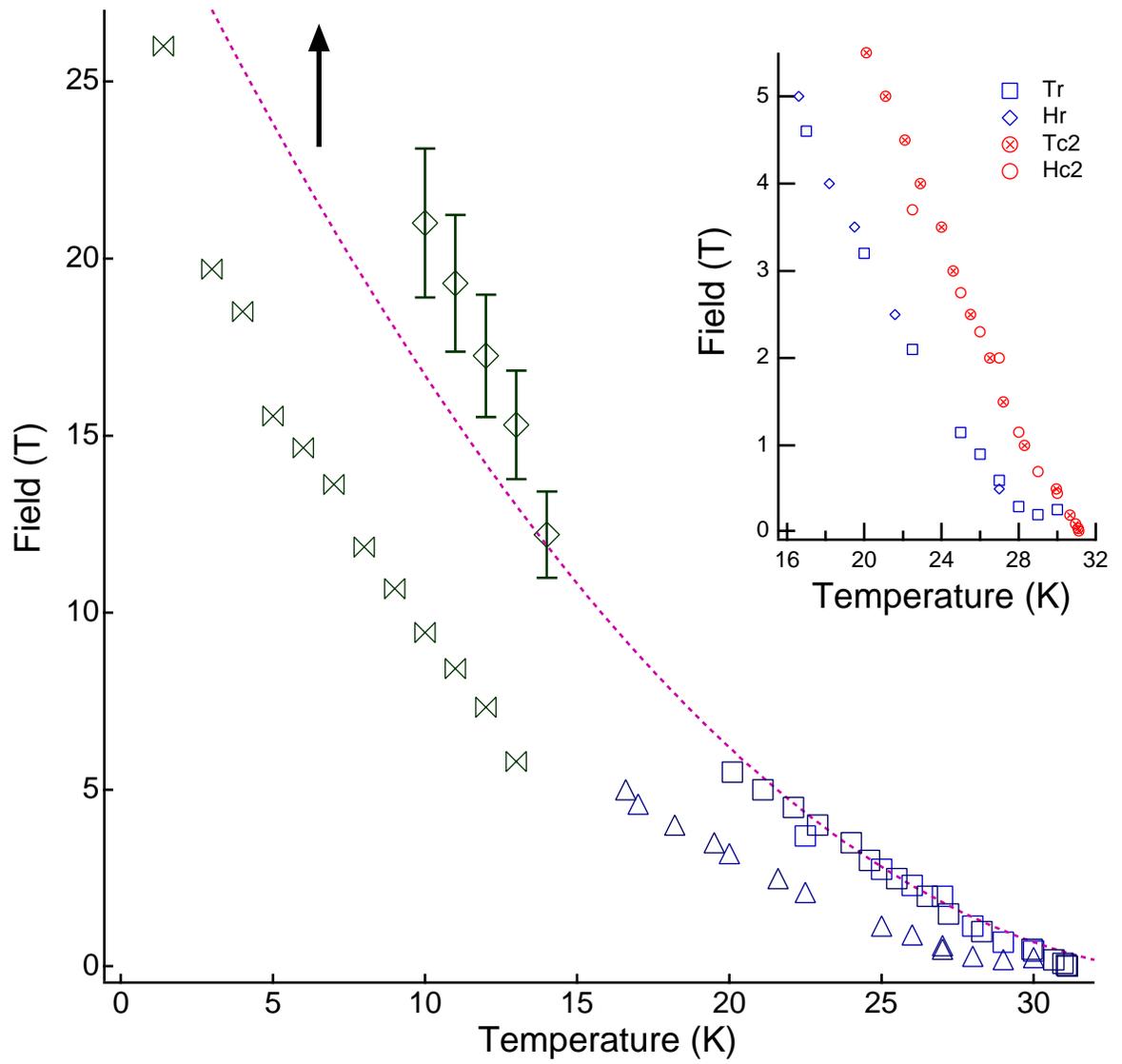
\begin{figure}[tbp]
	$$\epsffile{hc2_hr_inset.epsf}$$
	\caption{}
\label{hc2vst}
\end{figure}

Figure \ref{hc2vst}.
Values of H$_{c2}$\ measured as a function of temperature for
BKBO are shown along with the values of H$_{r}$\ and H$_{c2}$. 
H$_{c2}$\ values from both constant field and constant temperature
measurements are represented by a $\Box$ while  H$_{c2}$\
values for the sample used in the force magnetometer
measurements, are represented by a $\Diamond$.  The arrow indicates
that at temperatures below 10 K H$_{c2}$\ exceeded fields of 27 tesla, 
and 
H$_{r}$\ values from both constant field and constant temperature
measurements are represented by a $\triangle$.  The H$_{r}$\
values from the force measurements are represented by a $\Join$.  The dotted line is
the phenomenological fit to other data given by Gantmakher {\em et al.}
\protect \citep{Gantmakher1996}.  The inset shows the measured values of H$_{c2}$\ ($\circ$),T$_{c2}$ ($\otimes$), H$_{r}$%
\ ($\Diamond$), and T$_{r}$ ($\Box$) as a function of temperature.

\vspace{0.25cm}

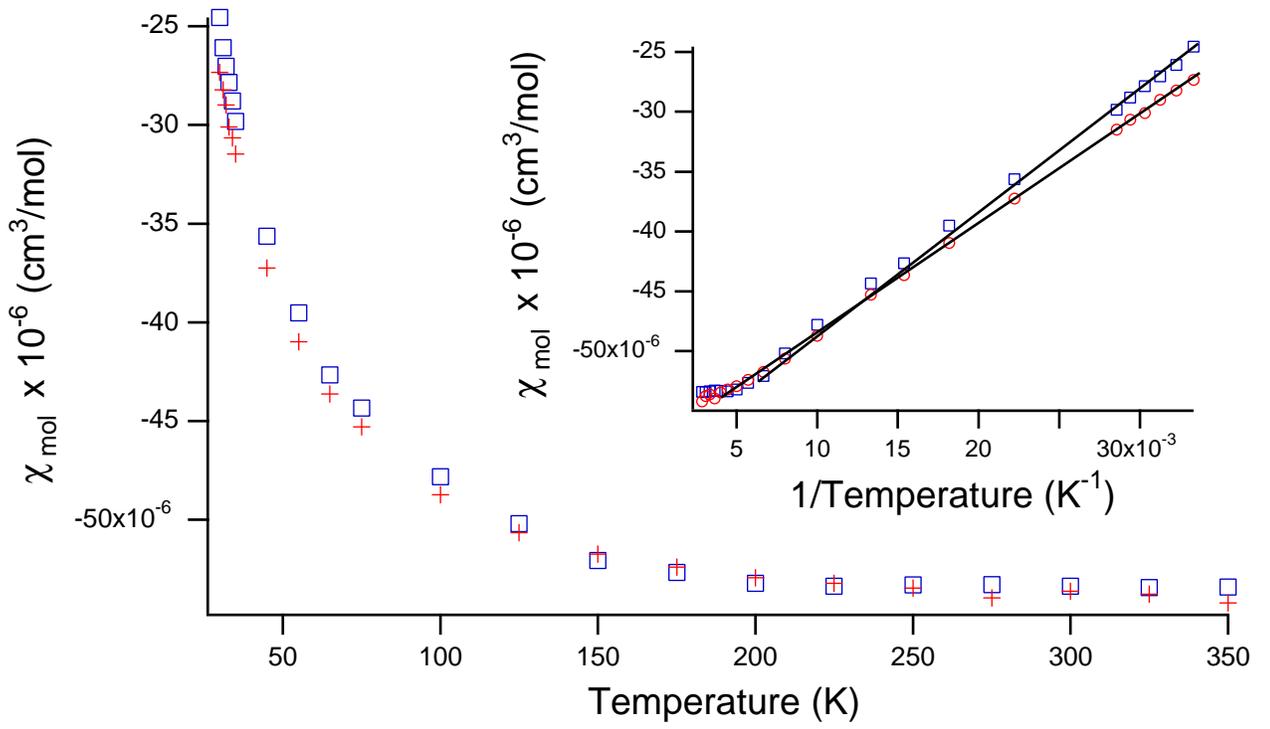
\begin{figure}[tbp]
	$$\epsffile{norm_state_suscept_papfig.epsf}$$
	\caption{}
\label{normmag}
\end{figure}

Figure \ref{normmag}.
The normal state susceptibility as a function of temperature for
two Ba$_{0.6}$K$_{0.4}$BiO$_{3}$\ samples.  The inset shows that 
$\chi$ is linear in 1/T for low temperatures.  The lines are a guide 
to the eye.

\vspace{0.25cm}

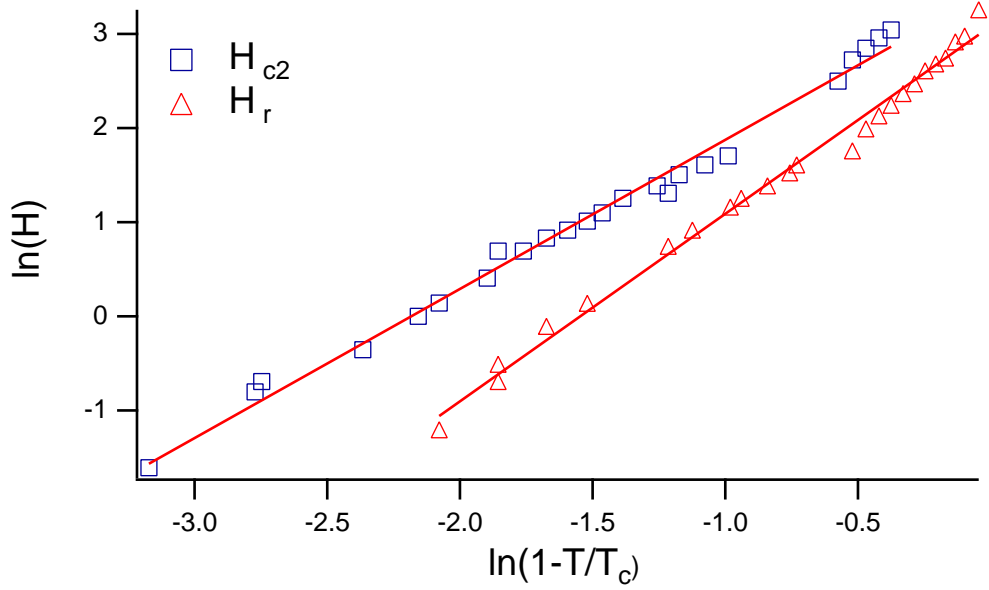
\begin{figure}[tbp]
	$$\epsffile{lnlnhc2hrvslnt.epsf}$$
	\caption{}
\label{fmlnlnhc2andhr}
\end{figure}

Figure \ref{fmlnlnhc2andhr}.
Temperature dependence of H$_{c2}$\ and H$_{r}$\ from the force
magnetometer and SQUID measurements combined.  Note that if only the high 
temperature SQUID data is used, the critical exponent drops to 1.25 for 
\Hctwo.

\vspace{0.25cm}

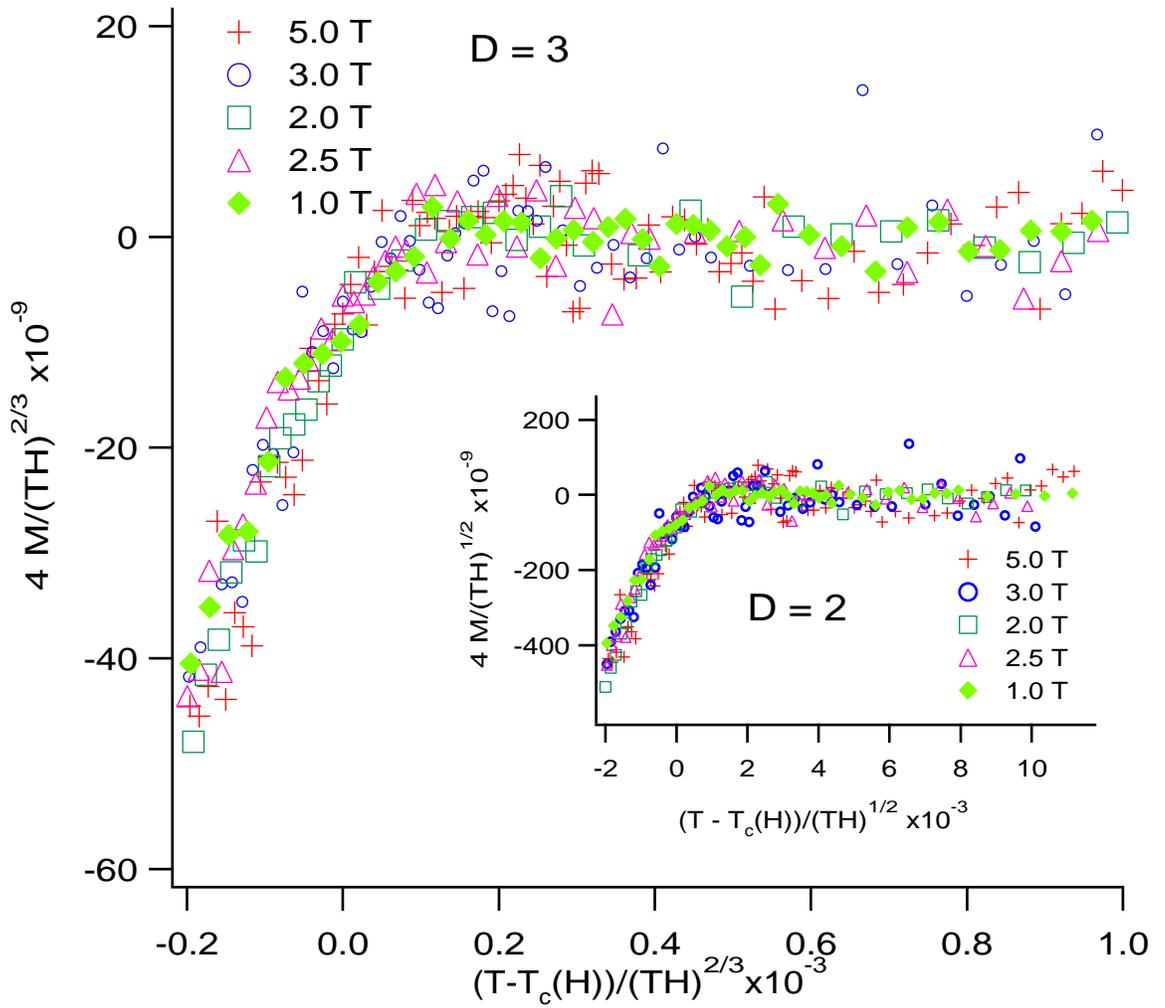
\begin{figure}[tbp]
	$$\epsffile{scaling.epsf}$$
	\caption{}
\label{scaling}
\end{figure}

Figure \ref{scaling}.
3D scaling of the magnetization data near \Tc.  2D scaling is 
shown in the insert.

\vspace{0.25cm}

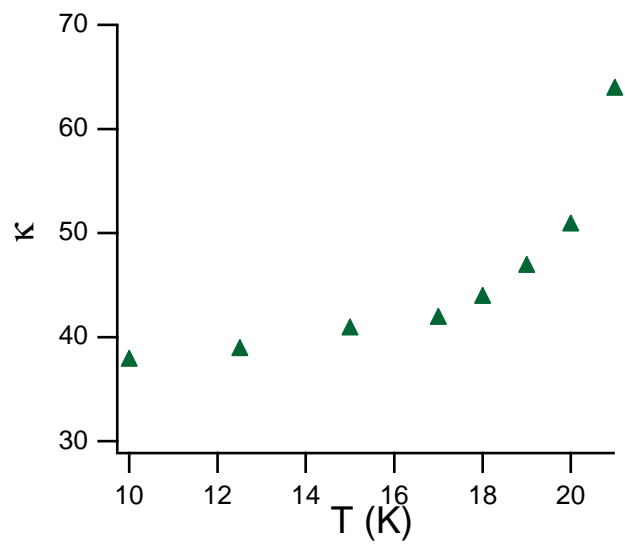
\begin{figure}[tbp]
	$$\epsffile{kappa.epsf}$$
	\caption{}
\label{kappavst}
\end{figure}

Figure \ref{kappavst}.
The Ginzburg-Landau parameter, 
$\kappa = \frac{\lambda}{\xi}$,  calculated from values of \Hcone\ 
and \Hctwo\ at selected temperatures as a function of 
temperature.

\newpage 

\newpage 

\newpage 

\newpage 

\newpage 

\newpage 

\newpage 

\newpage 

\newpage 

\newpage 

\newpage 

\end{document}